\documentclass[runningheads]{llncs}
\usepackage{graphicx}
\usepackage{acronym}
\usepackage{graphicx}
\usepackage{mathtools, amscd, mathrsfs}
\usepackage{multirow}
\usepackage{url}
\usepackage{tikz}
\usepackage{tikz-cd}

\usepackage{cite}
\usepackage{textcomp}
\usepackage{xcolor}

\usetikzlibrary{shapes.geometric}
\usepackage{amsfonts}
\usepackage{multirow}
\usepackage{colortbl}
\usepackage{bm}
\usepackage{array}

\usepackage{tikz}
\usetikzlibrary{arrows}

\def\eqref#1{equation~\ref{#1}}
\def\1{\bm{1}}

\DeclareMathAlphabet{\mathsfit}{\encodingdefault}{\sfdefault}{m}{sl}
\SetMathAlphabet{\mathsfit}{bold}{\encodingdefault}{\sfdefault}{bx}{n}

\newcommand{\ie}{i.e.,}
\newcommand{\eg}{e.g.,}

\definecolor{ballblue}{rgb}{0.13, 0.67, 0.8}
\definecolor{bananayellow}{rgb}{1.0, 0.88, 0.21}

\allowdisplaybreaks

\usepackage{newclude}
\usepackage{dirtree}
\usepackage{booktabs}
\usetikzlibrary{backgrounds}
\usepackage{subfig}
\usepackage{fixltx2e}
\usepackage{caption}
\usepackage{subfig}
\usepackage{lscape}
\usepackage{longtable}

\begin{document}
\title{Towards Using Probabilistic Models to Design Software Systems with Inherent Uncertainty\
	%\thanks{Thank you NWO!}
}

\titlerunning{Modeling Uncertainty During Design}

\author{Alex Serban\inst{1, 2}\and
	Erik Poll\inst{1}\and
	Joost Visser\inst{3}}

\institute{Radboud University,\and
Software Improvement Group\and%
Leiden University\\%
The Netherlands\\
	\email{a.serban@cs.ru.nl}}

% \author{Anonymous authors}
% \institute{}
\maketitle

\begin{abstract}
The adoption of \ac{ML} components in software systems raises new engineering challenges.
In particular, the inherent uncertainty regarding functional suitability and the operation environment makes architecture evaluation and trade-off analysis difficult. 
We propose a software architecture evaluation method called \ac{MUDD} that explicitly models the uncertainty associated to \ac{ML} components and evaluates how it propagates through a system.
The method supports reasoning over how architectural patterns can mitigate uncertainty and enables comparison of different architectures focused on the interplay between \ac{ML} and classical software components.
While our approach is domain-agnostic and suitable for any system where uncertainty plays a central role, we demonstrate our approach using as example a perception system for autonomous driving.
% For this system we empirically demonstrate that a component-based design is over 10\% more resilient to uncertainty than an end-to-end design.
% Moreover, we bring empirically evidence that architecture design patterns can help to significantly decrease the uncertainty associated to \ac{ML} components.

\end{abstract}

\acrodef{MUDD}[MUDD]{Modeling Uncertainty During Design}
\acrodef{ML}[ML]{machine learning}
\acrodef{DL}[DL]{deep learning}
\acrodef{NN}[NN]{neural networks}
\acrodef{DNN}[DNN]{Deep Neural Networks}
\acrodef{SVM}[SVM]{Support Vector Machine}
\acrodef{RL}[RL]{Reinforcement learning}
\acrodef{BN}[BN]{Bayesian network}

\acrodef{E2E}[ETE]{end-to-end learning}
\acrodef{uav}[UAV]{unmanned aerial vehicles}

\acrodef{EU}[EU]{epistemic uncertainty}
\acrodef{SU}[SU]{stochastic uncertainty}

% abbreviations
\acrodef{wrt}[\emph{w.r.t}]{with respect to}
\acrodef{st}[\emph{s.t.}]{such that}

% attacks
\acrodef{fgsm}[FGS]{Fast Gradient Sign}
\acrodef{bim}[BI]{Basic Iterative}
\acrodef{illcm}[ILC]{Iterative Least-likely Class}
\acrodef{jsma}[JSMA]{Jacobian-based Saliency Map Attack}
\acrodef{uap}[UAP]{Universal Adversarial Perturbations}
\acrodef{opa}[OPA]{One Pixel Attack}
\acrodef{pgd}[PGD]{projected gradient descent}
\acrodef{rssa}[RSSA]{Randomised Single Step Attack}
\acrodef{eat}[EAT]{Ensemble Adversarial Training}
\acrodef{gaa}[GAA]{Generative Adversarial Attacks}
\acrodef{gan}[GAN]{Generative Adversarial Networks}
\acrodef{nae}[NAE]{Natural Adversarial Examples}
\acrodef{atn}[ATN]{Adversarial Transformation Networks}
\acrodef{vae}[VAE]{Variational Auto-Encoders}
\acrodef{cfoa}[CFOA]{Complete First Order Adversary}
\acrodef{iid}[i.i.d]{independent and identically distributed}
\acrodef{bpda}[BPDA]{Backward Pass Differentiable Approximation}
\acrodef{alp}[ALP]{Adversarial Logit Pairing}
\acrodef{fbgan}[FB-GAN]{Featurized Bidirectional Generative Adversarial Networks}
\acrodef{sap}[SAP]{Stochastic Activation Pruning}
\acrodef{mat}[MAT]{Multi-strength Adversarial Training}
\acrodef{dam}[DAM]{Dense Associative Memory}
\acrodef{zoo}[ZOO]{Zeroth Order optimisation}
\acrodef{sa}[STA]{Strong Adversary}
\acrodef{lm}[LM]{Linear Models}
\acrodef{dt}[DT]{Decision Trees}
\acrodef{knn}[KNN]{K-nearest Neighbour}

\section{Introduction}
\label{sec:intro}

With the emergent adoption of \ac{ML} components in software systems, there is an increased need to tackle and harness their \emph{inherent} uncertainty.
Methods to address uncertainty exist for design time~\cite{meedeniya2012architecture,esfahani2013guidearch} and for run-time~\cite{esfahani2013uncertainty}.
However, previous work focused primarily on  uncertainty related to the parameters used to model a system or its context~\cite{esfahani2013uncertainty, meedeniya2012architecture,esfahani2013guidearch}.
\ac{ML} components add a new type of uncertainty that was only briefly explored previously; stemming from the fundamental impossibility to fully verify that they can satisfy their intended functionality and that they are able to cope with stochastic events during operation~\cite{serban2019designing}.

In this paper we introduce a method to evaluate architecture design alternatives for software using both traditional and \ac{ML} components.
The proposal, called \acf{MUDD}, is based on two guiding principles.
Firstly, the threats due to inherent uncertainty of \ac{ML} components are evaluated both locally (for the specific components) and tracked  as they propagate and influence other components in the system.
Secondly, the prior information about uncertainty of \ac{ML} components which is used at design time is considered incomplete and subject to continuous change.

The rest of the paper is organized as follows. 
Firstly, \ac{MUDD} is introduced (Section~\ref{sec:modeling}), followed by a demonstration (Section~\ref{sec:quantifying}), related work (Section~\ref{sec:backcground}) and conclusions (Section~\ref{sec:conclusions}).
\section{Modeling Uncertainty During Design (MUDD)}
\label{sec:modeling}

\ac{MUDD} explicitly models two sources of uncertainty specific to ``automated learning''~\cite{mahdavi2017classification}:
(1) epistemic uncertainty, \ie~the uncertainty about the data generation process (used for training \ac{ML} models), and (2) stochastic uncertainty, \ie~the uncertainty related to stochastic noise in the environment where a \ac{ML} component operates.
These uncertainty types have been studied in self-adaptive systems~\cite{perez2014uncertainties}, where software architecture plays an important role.
\ac{MUDD} is distinct by modeling these uncertainties at \emph{design} time, rather than at run time.

% \ac{MUDD} is a method to assess architecture design alternatives for software systems using both \ac{ML} and classical  components.
% In particular, the method focuses on reasoning about architectural design styles and patterns that can reduce the uncertainty stemming from defining software components and their interaction.
% \ac{MUDD} evaluates the impact of uncertainty in a bottom-up manner, starting locally, as it impacts specific \ac{ML} components, and tracked globally, as it influences other components in the system.
% This method allows fine-grained reasoning about design alternatives, where the changes between design alternatives can be evaluated at multiple levels.

Notably, \ac{MUDD} supports reasoning over which design alternatives are less sensitive to uncertainty and how design patterns can help mitigate it.
Moreover, the method allows to evaluate hypothetical scenarios, in which the data about uncertainty used at design time is incomplete or assumed to take any value.

From a methodological perspective, \ac{MUDD} only requires to annotate existing software architectures with the sources of uncertainty specific to \ac{ML} components.
Under the hood, \ac{MUDD} uses \acp{BN} to model a software system, propagate the uncertainties and obtain quantitative data about the architecture's sensitivity to uncertainty.

We emphasize that \ac{MUDD} uses these two uncertainty types
because they are application and context \emph{independent}, \ie~they are valid for any \ac{ML} model.
The methods used to measure them can be different, depending on the \ac{ML} algorithm employed. 
Therefore, they are parameters rather than fixed elements of \ac{MUDD}.
Nonetheless, \ac{MUDD} is not limited to any type of uncertainty.
% In fact, all uncertainty sources (application specific), can be modeled without any change.

Throughout the paper we use an example from autonomous driving, inspired by~\cite{behere2016functional,  serban2020standard}~--~the design of a perception system for scene understanding. 
The system performs three tasks: (1) object detection, which aims to identify the location of all objects in an image, (2) semantic segmentation, which assigns each pixel in an image to a predefined class, and (3) depth estimation, which determines the position of obstacles or the road surface.

The outcome of the example perception system is used in planning the next driving maneuvers.
The functionality of all components is implemented using \ac{DL} because no specification can be written for it, and other \ac{ML} algorithms perform worse.
We are interested to evaluate software architecture design alternatives and select the one which is the least sensitive to uncertainty.

\begin{figure*}[t]
	\centering
	\subfloat[End-to-end architecture.\label{fig:normal_end}]{\includegraphics[width=4cm, keepaspectratio]{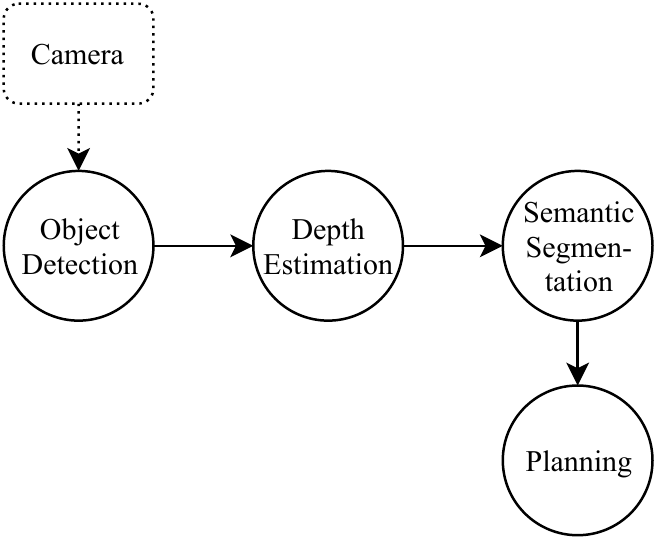}}
	\qquad\qquad\qquad\quad
	\subfloat[Component-based architecture.\label{fig:normal_dist}]{\includegraphics[width=4cm, keepaspectratio]{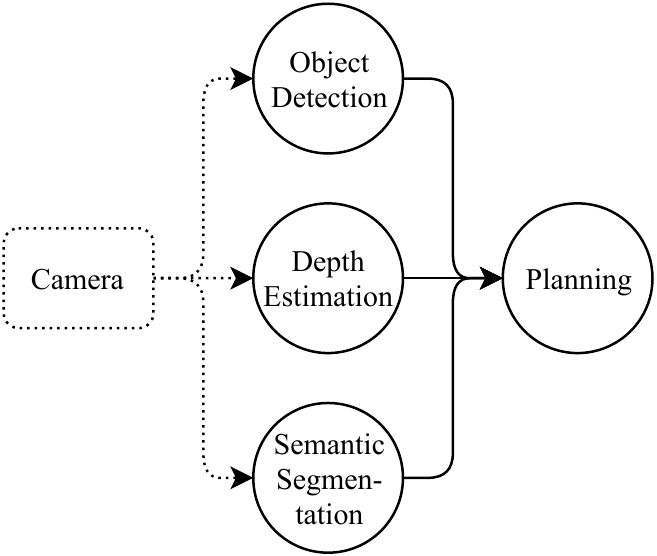}}
	\caption{Functional architectures for a scene understanding system in autonomous vehicles.}
	\label{fig:normal}
\end{figure*}

In Figure~\ref{fig:normal_end} and Figure~\ref{fig:normal_dist} we present two architecture candidates inspired by~\cite{serban2020standard} and~\cite{behere2016functional}.
The relevant functional components are illustrated using circles while the input coming from the camera is depicted with a rectangle.
The latter will not be later considered a node in the \ac{BN} (therefore its shape).

The first figure illustrates the end-to-end paradigm, where all components of the system are jointly trained to form a representation relevant to planning.
This corresponds to the recommendation in~\cite{serban2020standard}.
%, and adopted in~\cite{bojarski2016end}.
% The components are separated because they are trained using different objective functions and are subject to distinct drawbacks (also called multi-task learning in the \ac{ML} literature).
The components share a base network for feature extraction and have independent layers to decode the features for each task.
An alternative architecture is presented in the Figure~\ref{fig:normal_dist}, where the system is organized into distinct \ac{ML} components and integrated during planning.
This corresponds to the architecture recommended in~\cite{behere2016functional}.
We have chosen these architectural styles as the only alternatives we could find in literature.
However, \ac{MUDD} is not limited to any architectural style.
% In Section~\ref{sec:patterns} we discuss the adoption of architectural design patterns to reduce uncertainty.

\begin{figure*}[t]
	\centering
	\subfloat[End-to-end annotated architecture, as required by \ac{MUDD}.\label{fig:normal_unc_end}]{\includegraphics[width=4.5cm, keepaspectratio]{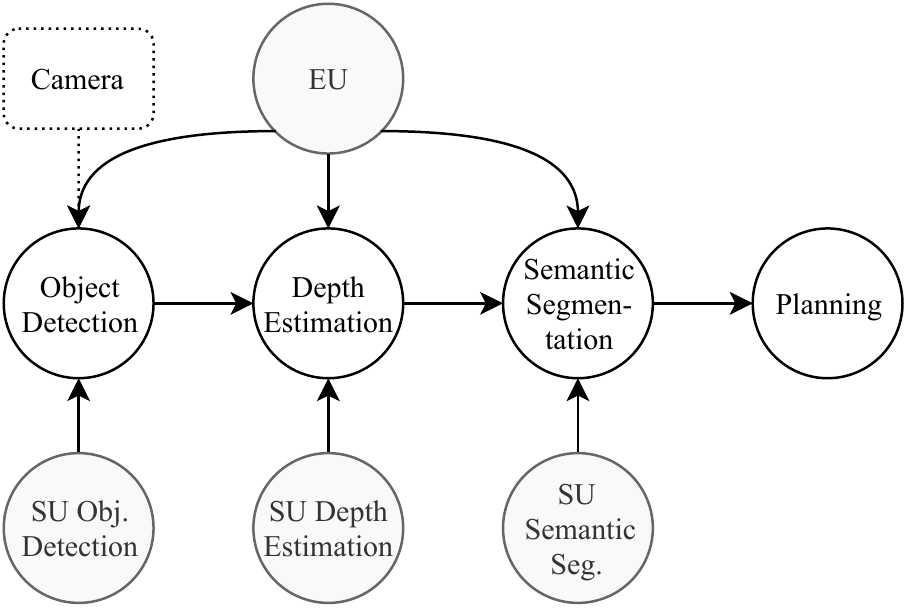}}
	\qquad\qquad\qquad\quad
	\subfloat[Component-based annotated architecture, as required by \ac{MUDD}. \label{fig:normal_unc_dist}]{\includegraphics[width=4.5cm, keepaspectratio]{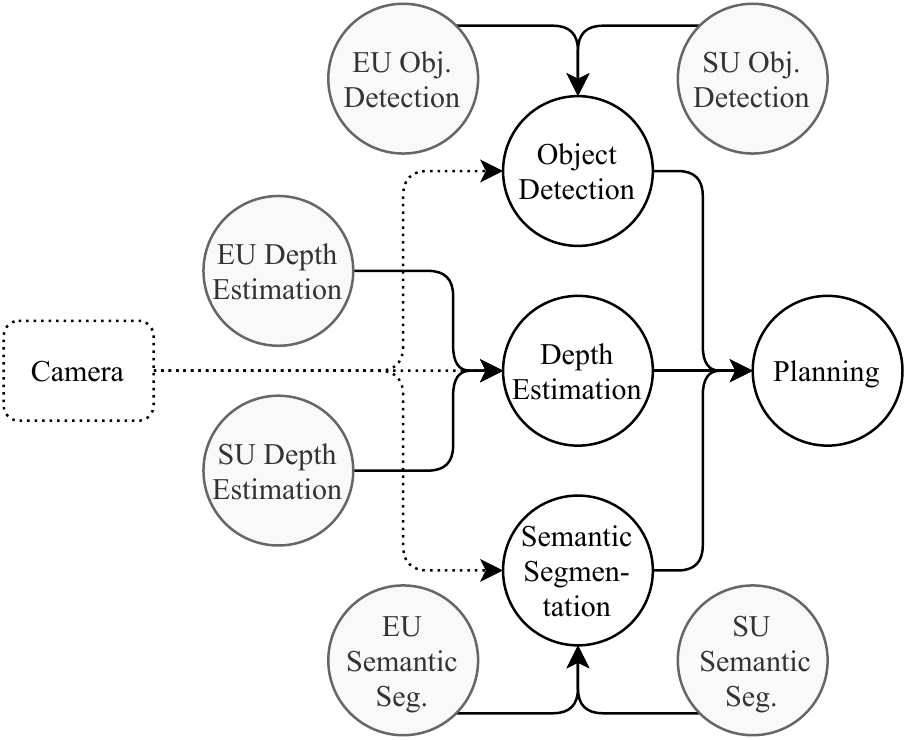}}
	\caption{Uncertainty representation for the two architectures presented in Figure~\ref{fig:normal}, where EU stands for epistemic uncertainty and SU for stochastic uncertainty.}
	\label{fig:normal_unc}
\end{figure*}

For reasoning about uncertainties, we propose to annotate the the two architectures with the sources of uncertainty specific to each component.
An example is given in Figure~\ref{fig:normal_unc}, which departs from the functional view presented in Figure~\ref{fig:normal} by illustrating the uncertainty sources, for each component.
In the first case, Figure~\ref{fig:normal_unc_end}, one base encoder is used for all tasks.
Therefore, only one node representing epistemic uncertainty (EU) influences all components.

% Arguably, the internal representation of the encoder might hold different representations for each task and be subject to distinct epistemic uncertainties.
% However, the internal representation is entangled and specific attributes corresponding to each task can not be easily extracted.

Different sources of stochastic uncertainty (SU) can impact the three tasks because one random event in the operational environment can influence segmentation, but not detection or depth estimation (and vice versa).
Therefore, for each component there is a different variable for stochastic uncertainty.
In the second scenario from Figure~\ref{fig:normal_unc_dist}, the components process raw data from camera independently.
Therefore, they are subject to distinct epistemic and stochastic uncertainties.
We note that these decisions are not application and context specific.
All \ac{ML} components are subject to these types of uncertainty.
% More examples, outside \ac{DL} are referenced in Section~\ref{sec:conclusions}.

% The annotated architectures from Figure~\ref{fig:normal_unc} represent the topology of a \ac{BN} and can be used for qualitative analysis.
% For quantitative results, the topology needs to be enriched with probability data.
% \input{qualitative}
\section{Quantitative Architecture Evaluation}
\label{sec:quantifying}

Under the hood \ac{MUDD} uses \acp{BN} to process quantitative data about uncertainties.
The probabilities needed to populate the network can be defined by experts or inferred through simulations.
The random variables in the \ac{BN} can take continuous or discrete values.
In the former case, the system designer chooses an a priori distribution for each variable, before seeing any data, and updates its parameters once new observations are available.
In the latter case, the variables take discrete values and are described by their probability mass functions.

For simplicity, we choose to model all variables through probability mass functions with two discrete values: \emph{low} or \emph{high} uncertainty.
When the uncertainty is low, the system is likely to satisfy its intended functionality and vice versa.
% Deciding if the uncertainty values are discrete or continuous is application and context specific.
% If the uncertainty in the \ac{ML} components used in a system can be modeled as a probability distribution and if the parameters of the distribution can be estimated well, a continuous approach may be better suited.
% Nonetheless, interpreting the parameters of a distribution requires more knowledge and may complicate the decision process.
%
Given the two proposed values for uncertainty, we are interested in evaluating the influence of different nodes in the network on planning and obtain quantitative results for the qualitative evaluation presented earlier.
Both the probabilities and the thresholds can be decided by domain experts or by simulation.

For the running example we use a test data set to extract the uncertainty estimates from \ac{DL} components, by averaging over samples in this data set.
The thresholds between low and high represent the lowest uncertainty estimate from the incorrectly classified examples in the testing data set.
The probability that a component has high (epistemic or stochastic) uncertainty will be the total number of test examples which have uncertainty higher than the threshold over the total number of testing examples.
Note that the correctly classified examples with high uncertainty will contribute to the probability that a component has high uncertainty.
This choice is deliberate because the system we study is safety-critical and uncertain decisions should be avoided altogether.

The conditional probabilities~--~\ie~the influence of components to the connected components~--~are evaluated in a similar manner.
They represent the probabilities that a component has high uncertainty, given the uncertainty values of the parent variables.
For example, $P(\mathit{OD}=\text{H} | EP=\text{H}, \mathit{SU}=\text{H})$ is the probability that the object detector is highly uncertain when the model has high epistemic and high stochastic uncertainty.
We use the same method and data set as before, but average the results when the parent variables have the same value.
The thresholds are also chosen as before.

\textbf{Uncertainty estimation.} All experiments are carried out using the CityScape data set~\cite{cordts2016cityscapes}.
For the end-to-end architecture presented in Figure~\ref{fig:normal_end} we train a variant of MultiNet~\cite{teichmann2018multinet} using an encoder based on the  DenseNet architecture, pre-trained on the ImageNet data set with a dropout probability of $p=0.2$.
We use different loss functions in a multi-task learning setting for object detection, depth estimation and semantic segmentation.
Epistemic uncertainty is approximated by casting a Bernoulli distribution over the model's weights and sample it at evaluation time using the dropout layers in the base encoder.
The mean of the dropout samples is used for prediction and the variance to output the uncertainty for each class.
Stochastic uncertainty is extracted from the final layer of each task.
% The geometry of the scene is interpreted using depth estimations from the base encoder, as in~\cite{eigen2015predicting}, while object detection and semantic segmentation have the same loss functions as in~\cite{teichmann2018multinet}.
For the component-based architecture presented in Figure~\ref{fig:normal_dist} we use an independent encoder and decoder for each task. 
Training is performed by minimizing the task specific loss function used in the multi-task setting described above.
The implementation of \ac{DL} components was done in Pytorch\footnote{https://pytorch.org/} and the \acp{BN} in Pomegranate\footnote{https://github.com/jmschrei/pomegranate}.
The uncertainty estimates are presented in  Table~\ref{tbl:case1} for the system in Figure~\ref{fig:normal_end} and Table~\ref{tbl:case2} for the system in Figure~\ref{fig:normal_dist}.

The heuristics applied to populate the tables represent the prior knowledge we embed in the network.
Depending on the context, software designers may choose to embed more domain knowledge or rely on expert opinion.
% For example, in the context of autonomous driving we may choose to augment the testing set with common perturbations, specific to different seasons, driving conditions or even malicious perturbations~\cite{hendrycks2019benchmarking}.

Given the probability tables, we can use the inference rules of \acp{BN} to answer questions about the proposed architectures.
We provide a working example: \eg~we wish to get quantitative evidence about the impact of high stochastic uncertainty in depth estimation on planning.
Setting depth estimation stochastic uncertainty to "High" ($SU_{DE} = \text{H}$), we can compute the
final impact on planning as follows. 
Let $\pi(x)$ represent the parent variables of node $x$ (the nodes that have a directed edge to it).
The probability that planning will have high uncertainty is:
\begin{equation*}
 \begin{array}{l}
	P(Planning=\text{H}) = P(SS |~\pi (SS)) \cdot P(DE |~\pi(DE))\cdot P (OD |~\pi(OD))~\cdot \\ P(SU_{SS}) \cdot P(SU_{DE}=\text{H}) \cdot P(SU_{OD}) \cdot P(EU),
\end{array}
\end{equation*}
for the end-to-end architecture and: 
\begin{equation*}
\begin{array}{l}
 P(Planning=\text{H}) = P(SS |~\pi (SS)) \cdot P(DE |~\pi (DE)) \cdot P (OD |~\pi (OD))~\cdot \\ P(SU_{SS}) \cdot  P(SU_{DE}=\text{H}) \cdot P(SU_{OD}) \cdot P(EU_{SS}) \cdot P(EU_{DE}) \cdot P(EU_{OD}), 
\end{array}
\end{equation*} 
for the component-based architecture, where the acronyms are as in Table~\ref{tbl:case1}~or~\ref{tbl:case2}.

\begin{table}[t!]
	\centering
	\subfloat{
		\begin{tabular}{c|c|c|c|c}
			$P (\cdot)$ & $EU$ & $SU_{OD}$ & $SU_{DE}$ & $SU_{SS}$\\
			\hline
			$H$ & 0.18 & 0.16 & 0.11 & 0.19 \\
%			$L$ & 0.5 & 0.5 & 0.5 & 0.5
		\end{tabular}}
	\qquad
	\subfloat{
		\begin{tabular}{c|c}
			$P (Planning$ & $SS)$ \\
			\hline
			0.1 & L  \\
			0.9 & H
	\end{tabular}}		
	\qquad\newline
	\subfloat{
	\begin{tabular}{c|c|c}
		$P (OD$ & $EU$ &$SU_{OD})$ \\
		\hline
		0.0 & L & L \\
		0.64 & L & H \\
		0.61& H & L \\
		1 & H & H
\end{tabular}}
\qquad
	\subfloat{
		\begin{tabular}{c|c|c|c}
			$P (DE$ & $EU$ &  $SU_{DE}$ & $OD)$ \\
			\hline
			0.0 & L & L & L  \\
			0.13 & L & L & H \\
			0.76 & L & H & L \\
			0.85 & L & H & H \\
			0.43 & H & L & L \\
			0.78 & H & L & H \\
			0.9 & H & H & L \\
			1 & H & H & H 	
		\end{tabular}}
	\qquad
	\subfloat{
		\begin{tabular}{c|c|c|c}
			$P (SS$ & $ EU$ & $SU_{SS}$ & $DE)$ \\
			\hline
			0.0 & L & L & L  \\
			0.28 & L & L & H \\
			0.64 & L & H & L \\
			0.72 & L & H & H \\
			0.66 & H & L & L \\
			0.58 & H & L & H \\
			0.61 & H & H & L \\
			1 & H & H & H 	
		\end{tabular}}		

	\caption{Independent and conditional probabilities for the end-to-end architecture in Figure~\ref{fig:normal_unc_end}. The acronyms used are OD~--~object detection, DE~--~depth estimation, SS~--~semantic segmentation, EU~--~epistemic uncertainty and SU~--~stochastic uncertainty. The uncertainty values are L~-~low and H~-~high.}
	\label{tbl:case1}
\end{table}
\begin{table}[t!]
	\centering
	\subfloat{
		\begin{tabular}{c|c|c|c|c|c|c}
	$P (\cdot)$ & $EU_{OD}$ & $SU_{OD}$ & $EU_{DE}$ & $SU_{DE}$ & $EU_{SS}$ & $SU_{SS}$\\
	\hline
	H & 0.14 & 0.16 & 0.31 & 0.44 & 0.17 & 0.19 
	\end{tabular}}
	\qquad
	\subfloat{
		\begin{tabular}{c|c|c}
			$P (OD$ & $EU_{OD}$ &  $SU_{OD})$ \\
			\hline
			0.0 & L & L \\
			0.57 & L & H \\
			0.41 & H & L \\
			1.0 & H & H
	\end{tabular}}
	\qquad
	\subfloat{
		\begin{tabular}{c|c|c|c}
			$P (DE$ & $EU_{DE}$ &  $SU_{DE}$) \\
			\hline
			0.0 & L & L \\
			0.51 & L & H \\
			0.47 & H & L \\
			1 & H & H
	\end{tabular}}
\qquad %\newline
\subfloat{
	\begin{tabular}{c|c|c}
		$P (SS$ & $EU_{SS}$ &  $SU_{SS})$ \\
		\hline
				0.0 & L & L \\
				0.11 & L & H \\
				0.42 & H & L  \\
				1.0 & H & H
\end{tabular}}
\qquad
\subfloat{
	\begin{tabular}{c|c|c|c}
			$P (Planning$ & $SS$ &  $DE$ & $OD)$ \\ \hline
			0.0 & L & L & L  \\
			0.34 & L & L & H \\
			0.34 & L & H & L \\
			0.66 & L & H & H \\
			0.34 & H & L & L \\
			0.66 & H & L & H \\
			0.66 & H & H & L \\
			1 & H & H & H 
\end{tabular}}
	\caption{Independent and conditional probabilities for the component-based architecture  in Figure~\ref{fig:normal_unc_dist}. The acronyms used are described in Table's~\ref{tbl:case1} caption.}
	\label{tbl:case2}
\end{table}

\begin{figure*}[t]
	\centering
	\subfloat[Influence of high stochastic uncertainty in depth estimation on planning.\label{fig:plot_1}]{\includegraphics[width=5.9cm, keepaspectratio]{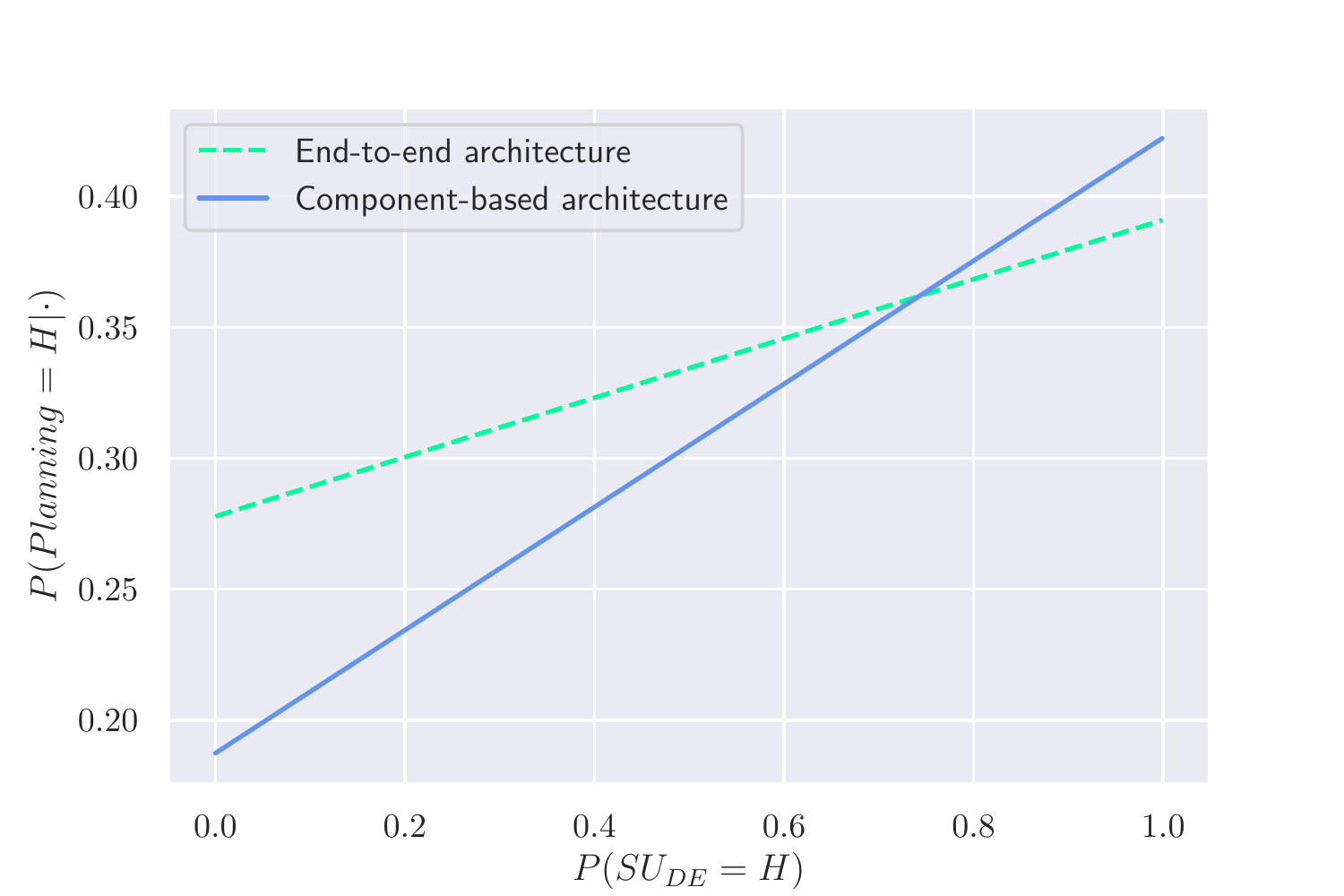}}
	\quad
	\subfloat[Influence of high stochastic uncertainty for depth estimation and all epistemic uncertainties on planning. \label{fig:plot_2}]{\includegraphics[width=5.9cm, keepaspectratio]{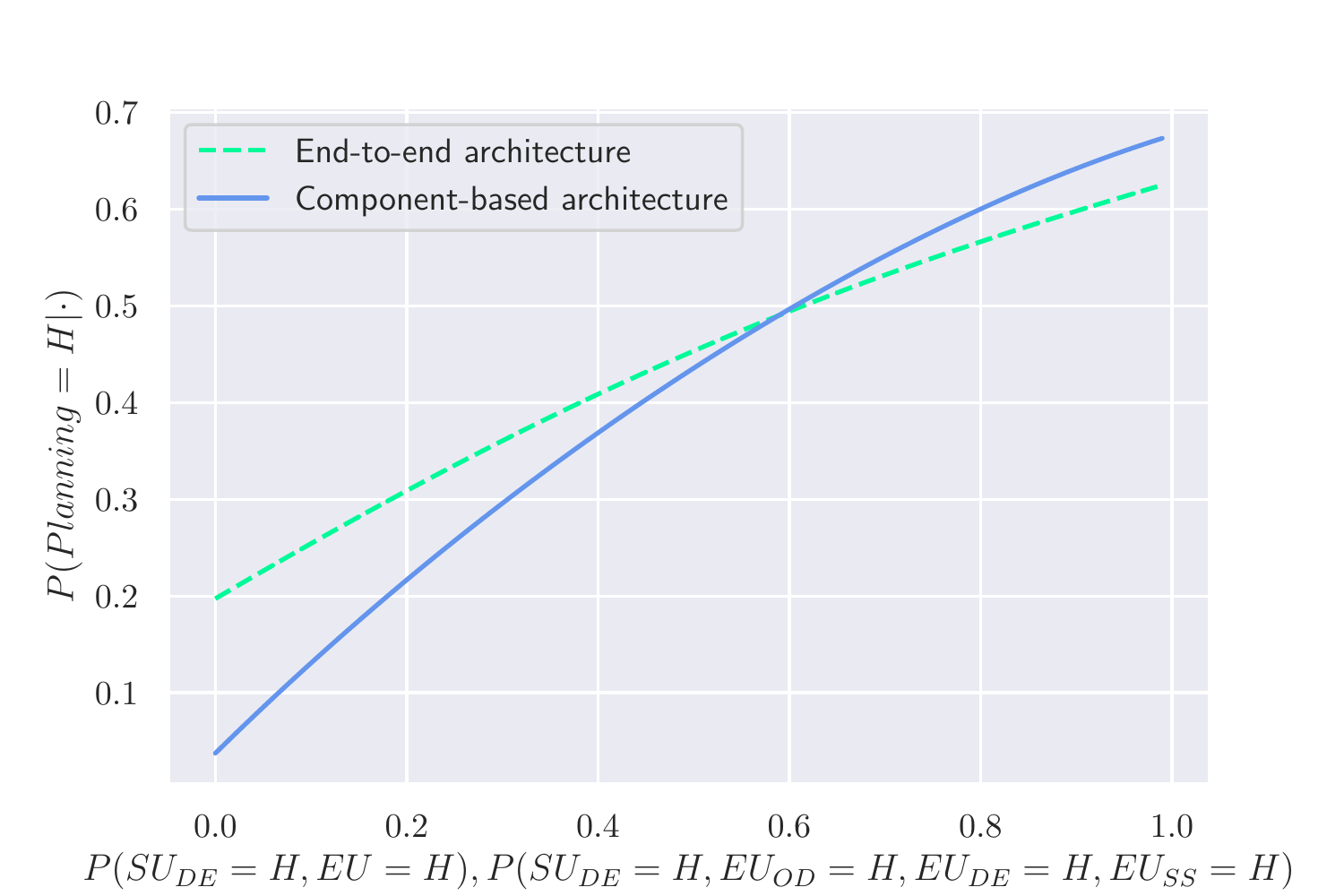}}
	\caption{Quantitative evaluation of uncertainty in software architecture design.}
	\label{fig:variance}
\end{figure*}

Running the computation we observe that the probability of uncertain planning is approximately $10\%$ lower for the component-based architecture (Figure~\ref{fig:normal_dist}) than for the end-to-end architecture.
Moreover, through the same model we can analyze how high stochastic uncertainty in depth estimation impacts planning within the minimum and maximum bounds.
We plot the probability that planning is uncertain given that depth estimation stochastic uncertainty is high, by varying $P(DE=\text{H}|SU=\text{H},~\cdot~)$ in Tables~\ref{tbl:case1} and~\ref{tbl:case2} between $[0, 1]$ with a step size of $0.01$.
The results are illustrated in Figure~\ref{fig:plot_1}.

The plot represents the influence of high stochastic uncertainty on depth estimation and the way it propagates on planning.
We observe that in the component-based architecture stochastic uncertainty in depth estimation has a lower impact on planning than in the end-to-end architecture, for values up to $\sim 0.7$, after which the end-to-end architecture is more resilient to uncertainty.
Depending on the operational environment, a software architect can choose the design that better fits the expected conditions.
For example, if an autonomous vehicle operates in limited domains~--~\eg~inside a warehouse~--~where the probability of encountering  stochastic events is low,  the component-based architecture for the scene understanding system is more appropriate.

Using the same model we can evaluate the influence of multiple  sources of uncertainty on planning.
We use the realistic assumption that the CityScape data set does not approximate all driving scenarios and thus may introduce high epistemic uncertainties.
Therefore, we evaluate the influence of all epistemic uncertainty sources on planning in the scenario described above, where we assume high stochastic uncertainty in depth estimation.
We use the same method as above to evaluate the probability that planning will have high uncertainty while we vary all epistemic uncertainty nodes simultaneously with the stochastic uncertainty in depth estimation.
The uncertainties vary between $[0,1]$, with a step size of $0.01$. % simultaneously?
The results are plotted in Figure~\ref{fig:plot_2}.

As in the previous case, the end-to-end architecture is more resilient to high uncertainties, for all the components mentioned above. 
Moreover, the threshold where the end-to-end architecture becomes more resilient than the component-based architecture is lower.
% Nonetheless, the impact on planning remains high for values near the threshold, where both architectures behave similarly.
% With a $50\%$ chance to plan actions that may lead to unintended outcomes, the system may not be usable.
However, epistemic uncertainty can be removed using more training data, so the scenario in which epistemic uncertainty is low is more realistic.
In this case, the component-based architecture is more resilient to uncertainty than the end-to-end architecture.
% This result implies that the component-based architecture is less likely to lead to unintended outcomes.

% \input{arch_patterns}
%\input{discussion}
\section{Related Work}
\label{sec:backcground}

At design time, the uncertainty in the parameters used to model a software system has been taken into account for evaluating the reliability of software architectures using robust optimization~\cite{meedeniya2012architecture}, for comparing software architectures when the impact of architectural decisions can not be quantified, using fuzzy methods~\cite{esfahani2013guidearch} and for evaluating trade-offs specific to quality attributes such as performance, using sensitivity analysis~\cite{etxeberria2014performance}.
However, none of these methods take into account the uncertainty related to ``automated learning", as indicated by~\cite{mahdavi2017classification}.

At run-time, various sources of uncertainty can be mitigated through self-adaptation~\cite{esfahani2013uncertainty}.
While several methods for self-adaptation use a related formalism, we tackle the problem at design time, and \emph{not} at run-time, as in self adaptation.
Therefore, self-adaptation is complementary, and a method that can unify uncertainty at design and run time is an interesting direction for future research.
\section{Conclusions and Future Work}
\label{sec:conclusions}

We introduce \ac{MUDD}, a method to evaluate and compare architecture design alternatives for systems using \ac{ML} components. 
In particular, we propose to explicitly model the inherent uncertainty specific to \ac{ML} components at design time, and evaluate how it propagates and influences other components in a system.
The proposed information needed to quantify the uncertainty for each \ac{ML} component is well studied both in the software architecture and  in the \ac{ML} literature.
For modeling software systems, \ac{MUDD} uses \acfp{BN}.

For future work we propose to further validate  the sources of uncertainty with practitioners (\eg~through interviews), and to facilitate the use of \ac{MUDD} by developing or integrating  with appropriate tools. 
New scenarios, which can better exhibit the potential of \ac{MUDD} and new uncertainty sources (\eg~\cite{serbanadversarial}) are planned as well.
Also, \acp{BN} are directed graphs and do not allow loops.
Alternatives that can overcome this limitation are planned for future work.

\bibliography{bibliography}
\bibliographystyle{splncs04}

\end{document}